\documentclass[twocolumn,showpacs,preprintnumbers,amsmath,amssymb,nofootinbib]{revtex4}
\usepackage{graphicx}
\usepackage[english]{babel}
\usepackage{amsfonts}
 \usepackage{amssymb}
 \usepackage{graphicx}
 
\usepackage{color}

\begin{document}

\title{Vacuum decay in an interacting multiverse}

\author{S. Robles-P\'{e}rez $^{1,2} $, A. Alonso-Serrano$^{1,2,3}$, C. Bastos$^{4}$, and O. Bertolami$^{5}$,  }

\affiliation{$^1$ Centro de F\'{\i}sica ``Miguel Catal\'{a}n'', Instituto de  F\'{\i}sica Fundamental, Consejo Superior de Investigaciones Cient\'{\i}ficas, Serrano 121, 28006
Madrid, Spain \\
 $^2$ Estaci\'{o}n Ecol\'{o}gica de Biocosmolog\'{\i}a, Pedro de Alvarado, 14, 06411 Medell\'{\i}n, Spain\\
 $^3$ School of Mathematics and Statistics, Victoria University of Wellington, PO Box 600, Wellington 6140, New Zealand\\
$^4$ GoLP/ Instituto de Plasmas e Fus\~ao Nuclear, Instituto Superior T\'ecnico, Avenida Rovisco Pais 1, 1049-001 Lisboa, Portugal \\
 $^5$ Departamento de F\'isica e Astronomia and Centro de F\'isica do Porto, Faculdade de Ci\^ encias da Universidade do Porto, Rua do Campo Alegre 687, 4169-007 Porto, Portugal }

\date{\today}

\begin{abstract}
We examine a new multiverse scenario in which the component universes interact. We focus our attention to the process of ``true" vacuum nucleation in the false vacuum within one single element of the multiverse. It is shown that the interactions lead to a collective behaviour that might lead, under specific conditions, to a pre-inflationary phase and ensued distinguishable imprints in the comic microwave background radiation.
\end{abstract}

\maketitle

\section{Introduction}

The idea that our universe is an element in a vast set of universes, the multiverse, has been argued to be an interesting way to address the cosmological constant problem in the context of string theory \cite{Susskind, Bousso, Weinberg}. Of course, this scenario raises many questions. How is the vacuum of our world chosen? Through anthropic arguments \cite{Polchinski}? Through quantum cosmology arguments \cite{Holman}? Is the string landscape scenario compatible with predictability \cite{Ellis}? Do the universes of the multiverse interact \cite{Bertolami2008} (see also Ref. \cite{Linde})? Does the multiverse exhibit collective behavior \cite{Alonso2012}?
The multiverse also arises in the context of the so-called Many World Interpretation of quantum mechanics \cite{Everett} and in the eternal inflationary model \cite{Linde1986}.

Actually, it has been recently proposed that the multiverse of eternal inflation and the many-worlds interpretation of quantum mechanics can be identified, yielding a new view on the measure and measurement problems \cite{Bousso&Susskind, Nomura}. However, it has been argued that a non-linear evolution of observables in the quantum multiverse would be an obstacle for such a description as these non-linearities are expected from quite general arguments \cite{Bertolami2013}.

In this paper we shall study the process of vacuum decay in the context of an interacting multiverse \cite{Bertolami2008, Alonso2012}. The consideration of an interacting multiverse entails a new and richer structure for the whole set of universes. The aim of this paper is to analize the influence of this enriched structure in the process of the vacuum decay of a single universe. First, we shall consider the Wheeler-De Witt equation for the wave function of the space-time. For many cases of interest the space-time is described by a homogeneous and isotropic geometry whose spatial section volumes scale as $a^3(t)$, where the scale factor $a(t)$ is a function of the cosmic time $t$ of a given multiverse. In this case the wave function of the universe, $\phi$, simplifies and it only depends on the values of the scale factor and the matter fields, i.e. $\phi = \phi(a,\vec{\varphi})$, with $\vec{\varphi} \equiv (\varphi_1(t), \varphi_2(t),\ldots)$ being a set of scalar fields. The
 se can t
 hus be considered as a field that propagates in the space spanned by the variables $\{a,\vec{\varphi}\}$.

Following the usual prescriptions of quantum mechanics, a second quantization procedure can be applied to the field $\phi(a,\vec{\varphi})$, which can be described in terms of quantum oscillators with their corresponding creation and annihilation operators. These operators would represent, in an appropriate representation, the creation and annihilation of pieces of the space-time with a given geometry. This description allows for representing the fluctuations of the space-time in terms of baby universes \cite{Strominger1990}, i.e. small particle-like portions of space-time that pop up and branch off from the parent space-time and propagate therein. Similarly, for a super-observer the field $\phi(a,\vec{\varphi})$ can be described in terms of particle-like pieces of space-time that we call universes.

The aim of this work is to examine if a supra-universal structure can influence the properties of a single causally isolated region of the space-time.

Whatever the definition of a universe is, it can be associate to some notion of causal closure, i.e. a region of the space-time manifold where all causally related events are self-contained. In other words, something that may cause or may be caused by any effect on any observed part of the universe should be included as being part of the universe. Thus, although it seems meaningless to consider \emph{external} elements of the universe, we shall see that this is not the case.

The classical and local notion of causal closure does not exclude the possibility that  non-local interactions among different regions of the space-time may determine some of the global properties of single universes. In fact, it has already been shown \cite{Alonso2012} that the interaction between two or more universes could determine the effective values of the cosmological constant of the universes. Despite that, light ray cones and local causal relations and properties within each single universe still obey the usual relations and remain causal. However, the value of a global property like its the cosmological constant can be affected by the interaction among the universes.

This cosmological picture is then completely different than the one single universe picture. Interactions and collective behaviour might then occur among the universes of the multiverse. Actually, this collective behaviour is fairly general and is at the very heart of quantum theory, which is a non-local theory and within which all the physical elements are fundamentally coupled to their environment and individual properties arise out of a result of some decoherence process. Thus, the true quantum state of the space-time must account for the states of all the universes, if they exist. The aim of this paper is to examine whether some of these collective processes may have an observable influence on the properties of our universe.

Irrespective of the consideration of a multiverse and its implications, it seems therefore interesting to analyze the influence, if any, that different distant regions of the space-time may have on the properties of the observable part of our local universe (see also Ref. \cite{Alonso-Serrano:2014dsa}) with a two-fold aim: i) to analyze whether they might help to solve some of the open questions posed by the latest Planck data \cite{Ade2014, Ijjas2013}, and ii) to look for distinguishable imprints of other universes in, for instance, the properties of the cosmic microwave brackground (CMB) spectrum \cite{Aguirre:2007an}.

This paper is organized as follows: In section II, we discuss the Hamiltonian quantum cosmology model of an interacting multiverse. In section III, we consider the bubble formation, that is, the nucleation of universes in a parent space-time and specifically address this nucleation in a setting where the universes are interacting. Finally, in section IV, we present a discussion of our results.

\section{The interacting multiverse}

Let us consider a simply connected piece of a homogeneous and isotropic space-time manifold endowed with a scalar field $\varphi$ that represents the matter content. More general topologies can also be considered by splitting the whole manifold into simply connected pieces of space-time \cite{Hawking1988}, each of which is quantum mechanically described by a wave function $\phi = \phi(a,\varphi)$ that is the solution of the Wheeler-DeWitt equation \cite{Alonso2012}
\begin{equation}\label{WDW01}
\ddot{\phi} + \frac{1}{a} \dot{\phi} - \frac{1}{a^2} \phi'' + \omega^2(a,\varphi) \phi = 0 ,
\end{equation}
where the scalar field has been rescaled according to Ref. \cite{Garay2013}, $\varphi \rightarrow \frac{2}{M_P}\sqrt{\frac{\pi}{3}} \varphi$, where $M_P$ is the Planck mass. In Eq. (\ref{WDW01}) the dots represent derivatives with respect to the scale factor and the prime denotes derivative with respect to the scalar field. The function $\omega(a,\varphi)$ contains the potential terms of the Wheeler-De Witt equation. In the case of a closed  space-time it is given by
\begin{equation}\label{FRQ01}
 \omega^2(a,\varphi) \equiv \sigma^2 (H^2 a^4 - a^2 ) ,
\end{equation}
where $\sigma \equiv \frac{3 \pi M_P^2}{2}$ and $H\equiv H(\varphi)$ is the Hubble function. The frequency $\omega$ has units of mass or, equivalently, units of the inverse of time or length. We shall consider two contributions to the Hubble function, i.e., $H^2 = H_0^2 + H_1^2$. The first one is due to the existence of a cosmological constant, $H_0^2={\Lambda_0 \over {3 M_P^2}}$, which is assumed to be very small. The second contribution is due to the potential of the scalar field, $H_1^2 = \frac{8 \pi}{3 M_P^2} V(\varphi)$.

Let us now develop a quantum field theory for the wave function $\phi$ in the curved minisuperspace spanned by $(a, \varphi)$ with a minisuperspace metric given by
\begin{equation}\label{MSM}
 G_{M N} = \left( \begin{array}{cc}
-a & 0  \\
0 & a^3 \end{array} \right) ,
\end{equation}
where $M, N$ stands for $\{a,\varphi\}$. The \emph{line} element of the minisuperspace metric is therefore
\begin{equation}\label{LE}
  d\mathfrak{s}^2 = - a d a^2 + a^3 d\varphi^2 .
\end{equation}
The scale factor, $a$, formally plays the role of the time variable  and the matter field the role of the spatial variable in the two dimensional Lorentzian minisuperspace metric (\ref{MSM})  ($a(t)$ can actually be seen as  a time reparametrization). We can now follow the usual procedure of a quantum field theory for the scalar field\footnote{Spinorial and vector fields could also been considered.} $\phi(a,\varphi)$ by considering the following action
\begin{equation}\label{S01}
 S = \int da d\varphi \, \, \mathcal{L}(\phi, \dot{\phi}, \phi') ,
\end{equation}
where the Lagrangian density is given, as usual, by
\begin{eqnarray}\label{L01}
  \mathcal{L} &=& \frac{1}{2} \sqrt{-G} \left\{ G^{M N} \partial_M \phi \partial_N \phi - \mathcal{V}(\phi)\right\} \\
   &=& \frac{1}{2} \left( - a \dot{\phi}^2 + \frac{1}{a} \phi'^2 \right) + \frac{a \, \omega^2}{2}  \phi^2 ,
\end{eqnarray}
where $G = {\rm det}(G_{M N})$. Then, the corresponding Euler-Lagrange equation \cite{Birrell1982}
\begin{equation}\label{EL01}
\frac{1}{\sqrt{-G}} \partial_M\left( \sqrt{-G} G^{M N} \partial_N \right) \phi + \frac{1}{2} \frac{\delta \mathcal{V}}{\delta \phi} = 0 ,
\end{equation}
turns out to be the Wheeler-De Witt equation, Eq. (\ref{WDW01}).

The Hamiltonian density that corresponds to the Lagrangian density, Eq. (\ref{L01}), is given by
\begin{equation}\label{H01}
 \mathcal{H} = -\frac{1}{2} \left( \frac{1}{a} P_\phi^2 + \frac{1}{a} \phi'^2 + a \omega^2 \phi^2 \right) ,
\end{equation}
where
\begin{equation}\label{MO01}
  P_\phi \equiv \frac{\delta \mathcal{L}}{\delta \dot{\phi}} = - a \dot{\phi}~,
\end{equation}
is the momentum conjugated to the field $\phi$.

We can now  pose an interaction scheme \cite{Bertolami2008, Alonso2012} among a set of $N$ universes by considering a total Hamiltonian density given by \cite{Alonso2012}
\begin{equation}\label{H02}
  \mathcal{H} = \sum_{n=1}^{N} \mathcal{H}_n^{(0)} + \mathcal{H}_n^{(i)} ,
\end{equation}
where $\mathcal{H}_n^{(0)}$ is the unperturbed Hamiltonian density of the $n$-universe, given by Eq. (\ref{H01}), and $\mathcal{H}_n^{(i)}$ is the Hamiltonian density of the interaction for the $n$-universe, that here we consider as the simple quadratic interaction between next neighbour universes,
\begin{equation}\label{H03}
 \mathcal{H}_n^{(i)} =  \frac{a \lambda^2(a)}{8} (\phi_{n+1} - \phi_n)^2 ,
\end{equation}
where $\lambda(a)$ is a coupling function that can depend on the value of the scale factor and we use periodic boundary conditions so that, $\phi_{N+1} \equiv \phi_1$.

We consider that the Hamiltonian density, Eq. (\ref{H02}), represents the evolution of a set of universes that are interacting to each other, where each internal observer do not see any interaction but only its own Hamiltonian density. We can take into account,  for the sake of simplicity, a new representation given in terms of the normal modes by means of the Fourier transformation of $\phi$ and $P_\phi$
\begin{eqnarray}
  \tilde{\phi}_k &=& \frac{1}{\sqrt{N}} \sum_{n=1}^{N} e^{-(2\pi i k n/N)} \phi_n , \\
  \tilde{P}_k &=& \frac{1}{\sqrt{N}} \sum_{n=1}^{N} e^{(2\pi i k n /N)} P_n ,
\end{eqnarray}
the Hamiltonian density, Eq. (\ref{H02}), becomes
\begin{equation}\label{H03}
  \mathcal{H} = -\frac{1}{2} \sum_{k=1}^{N}  \left( \frac{1}{a} \tilde{P}_k^2 + \frac{1}{a} \tilde{\phi}'^2_k + a \omega_k^2 \tilde{\phi}_k^2 \right) .
\end{equation}
The new quantum states oscillate now with a frequency given by
\begin{equation}\label{FRQ02}
  \omega_k^2(a,\varphi) = \omega^2(a,\varphi) +  \lambda^2(a) \sin^2\left( \frac{\pi k}{N}\right) .
\end{equation}
This is formally the typical quantum description of a collective system in terms of normal modes. These represent a collective behaviour in which the wave function can oscillate. For a single mode $k$ the oscillation of the wave function $\tilde{\phi}_k$ is given by the equation
\begin{equation}\label{WDW02}
  \ddot{\tilde{\phi}}_k + \frac{1}{a} \dot{\tilde{\phi}}_k - \frac{1}{a^2} \tilde{\phi}''_k + \omega_r^2(a,\varphi) \tilde{\phi}_k = 0 ,
\end{equation}
which is the effective Wheeler-De Witt equation of the wave function of the $k$-universe in the $\tilde{\phi}$ representation that appears as an isolated, non-interacting universe. We shall assume that, although we work in this single multiverse wave function formalism, all the results can be decomposed in terms of the previous formalism of individual universes.

The effective value of the potential term of the scalar field in the $r$-universe has been modified as a result of the interaction with other universes. Let us notice that Eq. (\ref{FRQ02}) can be written as
\begin{equation}\label{FRQ03}
\omega_k^2(a,\varphi) = \sigma^2 (\tilde{H}_{1,k}^2 a^4 + H_0^2 a^4 - a^2) ,
\end{equation}
where $H_0 = 3 \Lambda_0$ and $\tilde{H}_{1,k}^2 = \frac{8\pi}{3 M_P^2} \tilde{V}_k(\varphi, a)$, with
\begin{equation}\label{NP}
\tilde{V}_k(\varphi, a) = V(\varphi)  +  \frac{\lambda^2(a)}{4 \pi^2 a^4} \sin^2\left( \frac{\pi k}{N}\right) .
\end{equation}
Let us now analyse the influence of the last term in Eq. (\ref{NP}) in the terms of the $k$-universe. We restrict our interest to the regime where the wave function of the $k$-universe can be approximately described by the semiclassical wave funtion
\begin{equation}\label{WFSC01}
 \phi_k \approx e^{\pm \frac{i}{\hbar} S_0(a)} \Delta_k(a,\varphi) ,
\end{equation}
where $S_0$ is the action of the gravitational part alone with no interaction, that is:
\begin{equation}\label{CA01}
  S_0(a) = \sigma \int da \, a \sqrt{H_0^2 a^2 - 1} = \frac{\sigma}{3 H_0^2} ( H_0^2 a^2 - 1)^\frac{3}{2} ,
\end{equation}
where the positive and negative signs in Eq. (\ref{WFSC01}) correspond to the contracting and the expanding branches of $\phi_k$, respectively. The wave function $\Delta_k(a,\varphi)$ satisfies then, at first order in $\hbar$, the following wave equation \cite{Hartle1990}
\begin{equation}\label{SCEQ01}
  -i \hbar \frac{\partial}{\partial t} \Delta_k = \frac{1}{2} \left( \frac{1}{a} \frac{\partial^2}{\partial \varphi^2} + a \tilde{V}_k(\varphi,a) \right) \Delta_k .
\end{equation}
Let us notice that in the absence of any interaction scheme Eq. (\ref{SCEQ01}) is the Schr\"odinger equation for the scalar field $\varphi$ with a Hamiltonian given by
\begin{equation}\label{H04}
  h = \frac{1}{2} p_\varphi^2 + V(\varphi) .
\end{equation}
The field equation for the scalar field is given by
\begin{equation}\label{SF01}
  \ddot{\varphi} + 3\frac{\dot{a}}{a} \dot{\varphi} + \frac{d\tilde{V}_k}{d\varphi} = \ddot{\varphi} + 3\frac{\dot{a}}{a} \dot{\varphi} + \frac{dV}{d\varphi} = 0 ,
\end{equation}
where the dots stands for derivative with respect to the Friedmann time $t$, i.e. $\dot{\varphi} \equiv \frac{d \varphi}{d t}$. The last term in the potential $\tilde{V}_k$, given by Eq. (\ref{NP}), has no influence upon Eq. (\ref{SF01}), so the classical behaviour of the scalar field remains unaltered with respect the usual description. Quantum mechanically, however, the extra term in the potential introduces a modification in the vacuum state that has to be accounted for any vacuum decay process of the universe. This has a major influence in the process of bubble formation and in the global structure of the space-time.

\section{Bubble formation}

The quantum interactions among distant regions of the whole space-time manifold can modify the vacuum state of the matter fields. We have seen that for a chain of interacting universes, where the potential of the scalar field for a normal mode of the wave function of the universe depends on the value $r$ of the mode. Different universes may remain in different mode states and observers therein feel their universes to be filled with a scalar field whose vacuum state is different for each mode state. The universes may then suffer a process of vacuum decay between two different values of their vacua.
The key point for the vacuum decay is the existence of different local minima in the potential of the matter fields. The minimum of those local minima is called the true vacuum and the remaining are the false vacua. In the late 1970s Coleman has generalized the quantum mechanical tunnelling effect of transition from the ``false" vacuum (excited state) into the ``true" vacuum (ground state) in field theory \cite{Coleman1}. Subsequently, quantum radiative corrections were introduced \cite{Coleman2} and finally the effect of gravity was considered \cite{Coleman1980}. The issue of adopting a set of consistent boundary conditions for the decay process was examined in Ref. \cite{Mottola}. In the framework developed by Coleman and collaborators, a field in a false vacuum state can then decay to the state of true vacuum with a probability of occurrence per unit time per unit volume, $\Gamma/V$, given in the semiclassical approximation by
\begin{equation}\label{P01}
 \Gamma/V = A e^{-B/\hbar} ,
\end{equation}
where $A$ and $B$ are two quantities to be determined. The result is the materialization of a bubble of true vacuum separated by a thin wall from the surrounding false vacuum. The global  picture is then a vast space-time in the false vacuum splattered by bubbles of true vacuum that are continuously forming and expanding until they finally collide, merge and fill the whole space-time.

\begin{figure}[htbp]
\begin{center}
\includegraphics[width=9cm]{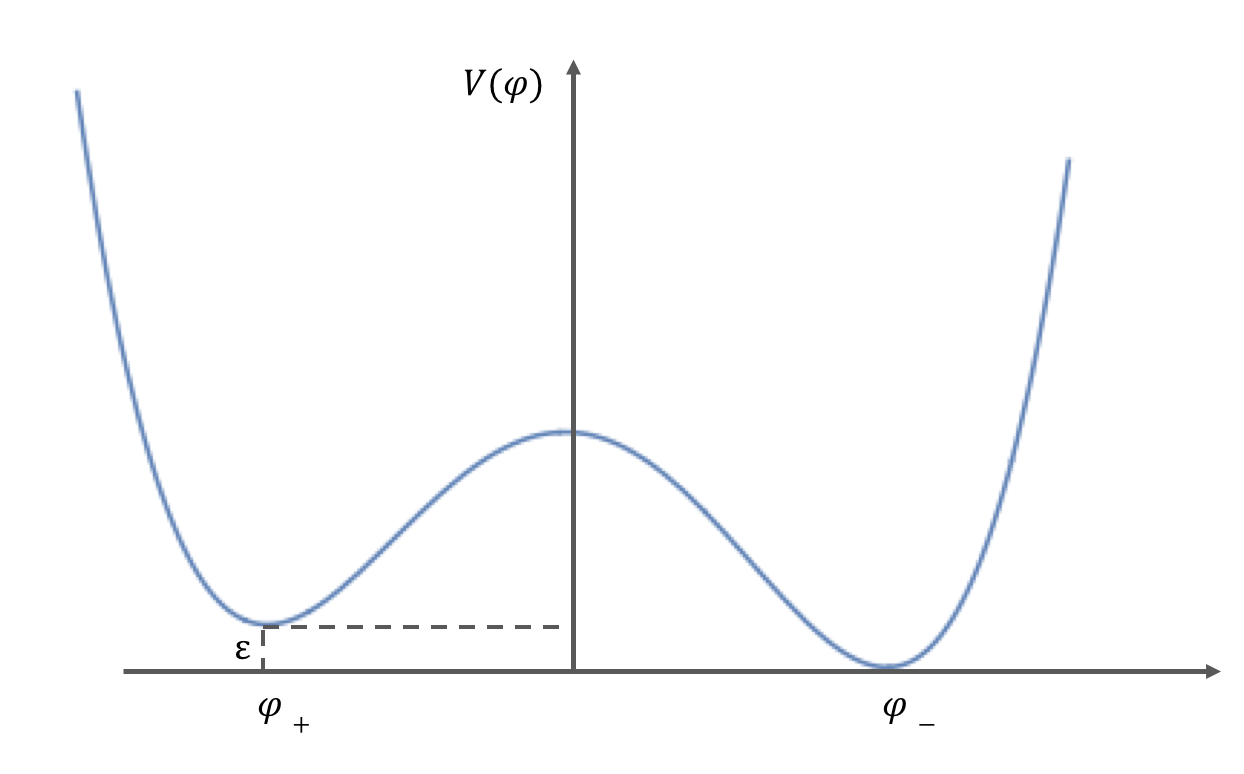}
\caption{Potential in Ref. \cite{Coleman1980}.}
\label{fig1}
\end{center}
\end{figure}

Let us now consider the process of vacuum decay in the context of a theory \cite{Coleman1980} with a potential $V(\varphi)$ that has two minima, $\varphi_+$ and $\varphi_-$, with  values of their vacuum energy given by $\Lambda_+ \equiv V(\varphi_+)$ and $\Lambda_- \equiv V(\varphi_-)$, respectively (see Fig. \ref{fig1}). The thin wall approximation, that is going to be applied, it is satisfied when
\begin{equation}\label{TW01}
  \Lambda_+ - \Lambda_- = \varepsilon \hspace{0.2cm}, \hspace{0.2cm} {\varepsilon\over \Lambda_{\pm}}\ll1 .
\end{equation}
Hence, a region of the space-time in the false vacuum may decay into the true vacuum by nucleating a bubble of true vacuum within the surrounding false vacuum that rapidly grows and expands when it is energetically favourable.

\begin{figure}[htbp]
\begin{center}
\includegraphics[width=9cm]{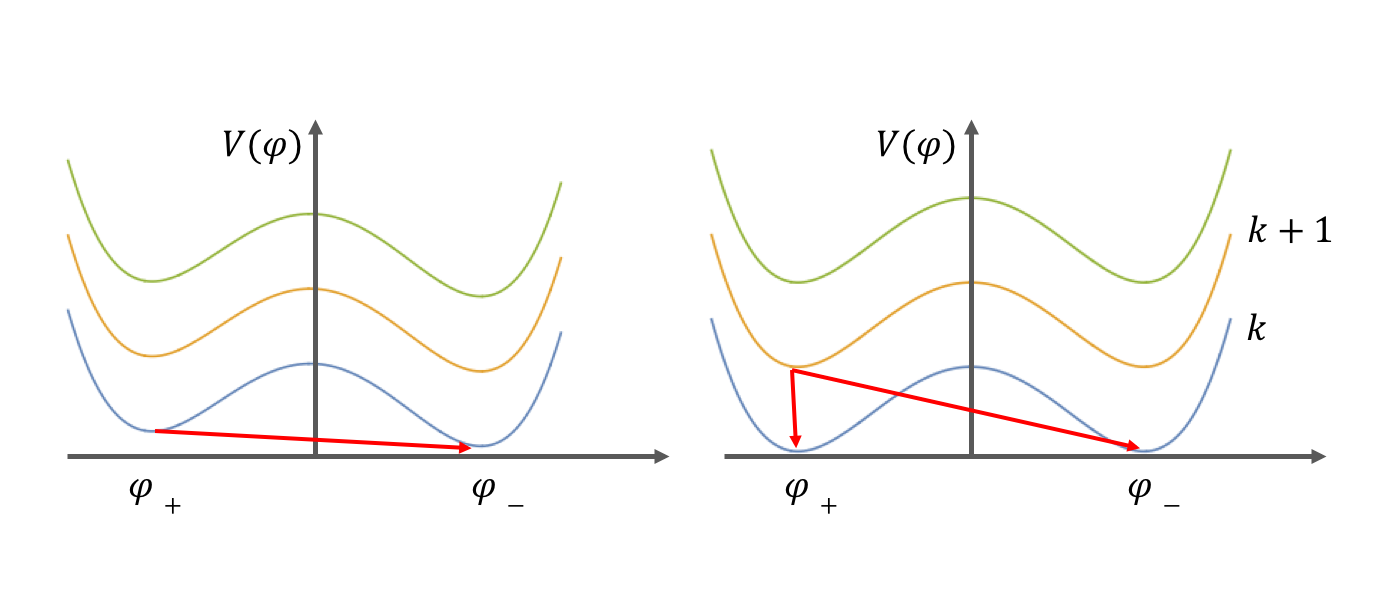}
\caption{a) Global picture of the potential (C-DL like potential); b) Quartic potential. Vacuum decay can now occur between two different minima of the same mode (C-DL) or between the minima of two different modes (interacting multiverse).}
\label{fig2}
\end{center}
\end{figure}

Let us now analyse the same process in the context of the interacting multiverse described in Sect. II. For a given value of the scale factor, the extra term of the potential of the $k$-universe is a constant that depends on the value $k$ of the normal mode of a given universe. The global effective value of the potential is then given by a set of curves separated by $k$ units (see Fig. \ref{fig2}), with $k = 0, \ldots, N/2$. The global picture presents then a landscape structure with $N$ different vacua: $N-1$ false vacua states and a true vacuum state $V(\varphi_+, k=0)$. We can consider, on one hand, the vacuum decay in a universe as a consequence of the multiverse interaction. On the other hand, we can also consider a global picture of vacuum in the multiverse and study the vacuum decay into a real vacuum that corresponds to a single universe.

The vacuum decay process follows the description of Ref. \cite{Coleman1980}. However, the process of bubble formation and the global structure of a single universe is now much richer. It can be envisaged as follows: small baby universes are created from quantum fluctuations of the space-time. At small values of the scale factor (i.e. values close to the Planck scale) the fluctuations of the scalar field and the effects of the interaction among the universes are dominant, so the newborn universes are expected to remain in normal modes with a high value of $k$. In some universes, the effective value of the potential would be high enough to trigger inflation even if we assume the new limit that is suggested by the most recent Planck data \cite{Ijjas2013}. As the universe expands in the $k$-false vacuum, different processes of vacuum decay are expected to occur generating new bubbles of smaller and smaller false vacua until the bubbles are created in the true vacuum $\Lambda_{+,k
 =0}$. Ho
 wever, the process does not stop there. The quantum fluctuations of the space-time of large regions with false or true vacuum state would supply new baby universes where the process of vacuum decay and bubble formation would take place continuously in a self-contained eternal process.

\begin{figure}[htbp]
\begin{center}
\includegraphics[width=8cm]{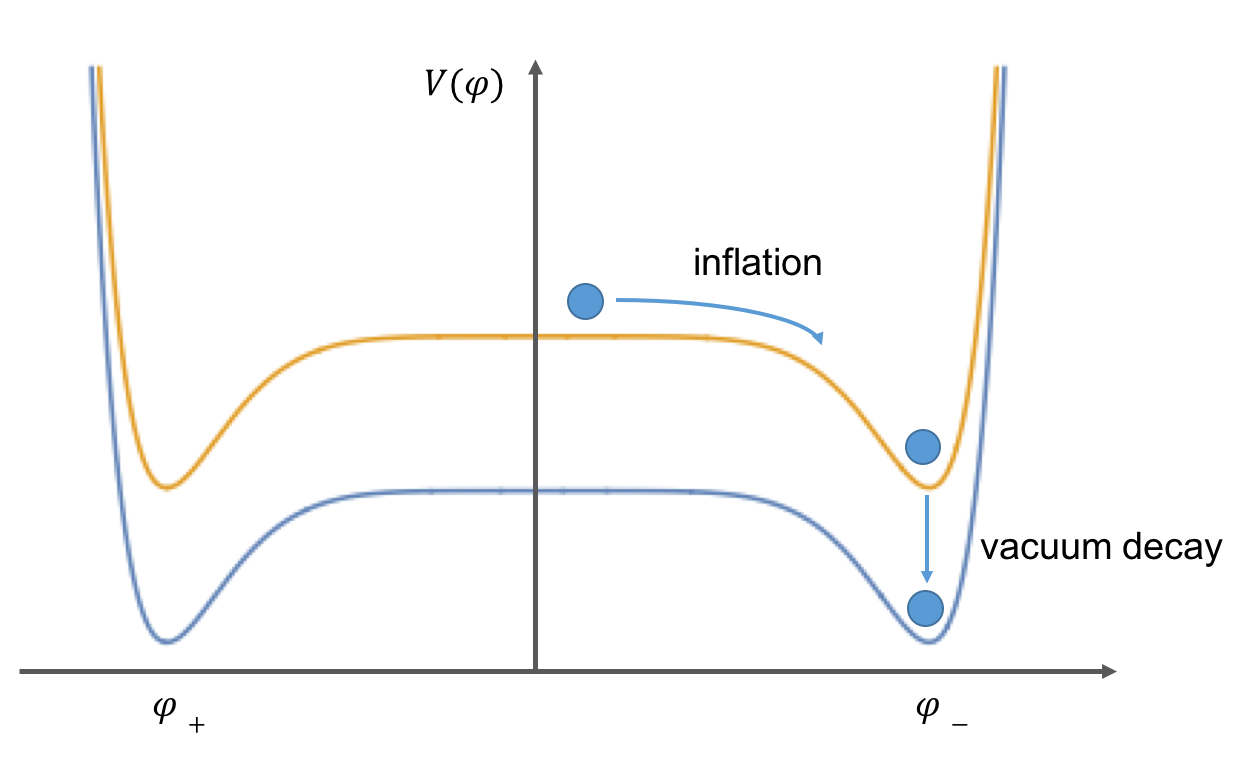}
\caption{Plateau-like potentials in the interacting multiverse. Even though the classical energy density of the plateau can be very small for a particular mode, i.e. $V_k(0) - V_k(\varphi_\pm) \ll M_P^4$, which is suggested by the most recent Planck data \cite{Ijjas2013}, the energy supplied by the quantum vacuum state could be large enough to trigger inflation in universes with a high mode state, i.e. $V_k(\varphi_\pm) - V_{k=0}(\varphi_\pm) \sim M_P^4$, for high values of $k$. A process of vacuum decay could then occur afterwards.}
\label{fig3}
\end{center}
\end{figure}

\subsection{Quartic potential}

For the sake of concreteness, let us consider the scalar field with a quartic potential
\begin{equation}\label{QV}
  V(\varphi) = \frac{m^4}{4 \lambda_{\varphi}^2}  - \frac{1}{2} m^2 \varphi^2 + \frac{\lambda_{\varphi}^2}{4} \varphi^4  ,
\end{equation}
where $m$ is the mass and $\lambda_{\varphi}$ is the self-coupling of the scalar field. It has two minima located at $\varphi_\pm = \pm \frac{m}{\lambda_{\varphi}}$, both with the same value of the potential given by $V(\varphi_\pm) = 0$. Let us notice that in the case studied by Coleman and De Luccia the value of the potential at the two different minima differs by a small amount of energy $\varepsilon$, otherwise the process of vacuum decay would not be possible. In the case under consideration, this condition is not necessary because the process of vacuum decay will take place between the vacuum states of different modes that could correspond to different universes or to a single universe whose vacuums had been modified by the interaction with other universes. This is schematically represented in Fig. \ref{fig2}b. Thus, instead of having Eq. (\ref{TW01}) defining the thin wall approximation, here we have 
\begin{equation}\label{TWint01}
\tilde{V}_k(\varphi_+)-\tilde{V}_{k-1}(\varphi_+)=\varepsilon_k~.
\end{equation}
That is 
\begin{equation}\label{TWint02}
\varepsilon_k= {\lambda^2(a)\over 4\pi^2 a^4} \left[\sin^2\left( \frac{\pi k}{N}\right)- \sin^2\left( \frac{\pi (k-1)}{N}\right)\right]~.
\end{equation}
The probability for a vacuum decay from the mode $k$ to the mode $k-1$ is given by \cite{Coleman1980}
\begin{equation}\label{B}
B = 2\pi^2 \bar{\rho}^3 S_1-{1\over 2} \pi^2 \bar{\rho}^4 \varepsilon_k~,
\end{equation}
where $\bar{\rho}=3S_1/\varepsilon_k$ and $S_1=2\int{d\rho [\tilde{V}_0(\varphi)-\tilde{V}_0(\varphi_+)]}$. $\tilde{V}_0(\varphi)$ is a function chosen such that $\tilde{V}_0(\varphi_+)=\tilde{V}_0(\varphi_{-})$ and that 
\begin{equation}\label{potential1}
{{d\tilde{V}_0}\over d\varphi}(\varphi_{\pm})=0~.
\end{equation}
Thus, $\tilde{V}_0(\varphi)=V(\varphi)$ given in Eq. (\ref{QV}). We choose $\bar{\rho}$ as the point at which $\varphi$ is the average of its two extreme values,
\begin{eqnarray}\label{phi12}
\phi_{1/2}&=&{1\over2}\left(\tilde{V}(\varphi_+, k)+\tilde{V}(\varphi_+, k-1)\right)\nonumber\\
&=&{\lambda^2(a)\over 4\pi^2 a^4} \left[\sin^2\left( \frac{\pi k}{N}\right)+ \sin^2\left( \frac{\pi (k-1)}{N}\right)\right].
\end{eqnarray}
Furhermore, $\bar{\rho}$ is assumed to be large compared to the length scale on which $\varphi$ varies significantly.
Then, it is possible to write $\varphi$ in terms of $\rho$ \cite{Coleman1980}:
\begin{eqnarray}\label{rho}
\rho-\bar{\rho}&=&\int_{\phi_{1/2}}^{\phi}{d\varphi [2 (\tilde{V}_0(\varphi)-\tilde{V}_0(\varphi_{\pm}))^2]^{-1/2}}\Leftrightarrow\nonumber\\
\Leftrightarrow \phi(\rho)&\!\!\!=\!\!\!&{m\over \lambda_{\varphi}} + \tanh\left[{{m\over 2\sqrt{2}}(\rho-\bar{\rho})+\tanh^{-1}{{\lambda_{\varphi}\over m}\phi_{1/2}}}\right]~.\nonumber \\
\end{eqnarray}
Hence, following Ref. \cite{Coleman1980} we now can evaluate $S_1$ in the thin wall approximation,
\begin{equation}\label{S}
S_1=2\int_{-\infty}^{+\infty}{d\rho [\tilde{V}_0(\varphi)-\tilde{V}_0(\varphi_+)]}={2\sqrt{2}\over3}{m^3\over \lambda_{\varphi}^2}~,
\end{equation}
and consequently the probability factor $B$ for a vacuum decay from the mode $k$ to the mode $k-1$ is
\begin{equation}\label{B2}
B = {10\over3}\pi^2 {m^{12}\over\lambda_{\varphi}^8}\left({{4\pi^2 a^4}\over \lambda^2(a)}\right)^3{1\over{ \left[ \sin^2\left( \frac{\pi k}{N}\right)- \sin^2\left( \frac{\pi (k-1)}{N}\right)\right]^3}}~.
\end{equation}

It is worth noticing that  Eq. (\ref{B2}) restricts the values of coupling function $\lambda(a)$ that suppress the vacuum decay at large values of the scale factor. For instance, with a polynomial value $\lambda(a) \propto a^n$, $n$ must satisfy $n \leq 2$ in order to fulfill the condition that the vacuum decay cannot grow with the scale factor. We can thus analyze some plausible cases.

\subsubsection{Case 1: $\lambda(a) \propto a^2$}

Let us first consider the value
\begin{equation}
	\lambda^2(a) = \frac{9 \pi M_P^2}{2} \Lambda a^4 ,
\end{equation}
where the constants has been chosen for later convenience. Then, Eq. (\ref{FRQ03}) can be re-written as
\begin{equation}\label{FRQ04}
	\omega_k^2(a,\varphi) = \sigma^2 (H_{0,k}^2  a^4  - a^2) + \sigma^2 H_1^2  a^4,
\end{equation}
where $H_1 = \frac{8 \pi}{3 M_P^2} V(\varphi)$, and
\begin{equation}
	H_{0,k}^2 =  3 \Lambda_k^{\rm eff} ,
\end{equation}
with
\begin{equation}\label{EFFL}
	\Lambda_k^{\rm eff} =  \Lambda_0 +  \Lambda \sin^2\frac{\pi k}{N} .
\end{equation}
Thus, Eq. (\ref{WDW02}) would represent the quantum state of a universe with an effective value of the cosmological constant of the background space-time given by Eq. (\ref{EFFL}). For a positive value of $\Lambda$,  $\Lambda^{\rm eff}_k \in [\Lambda_0, \Lambda_0 + \Lambda]$. If we assume a small value of $\Lambda_0$ (included the value $\Lambda_0=0$) and a value $\Lambda \sim M_P^4$, then, at the onset of the universe where it is supposed to remain at a large value for the $r$ mode, the effective value of the cosmological constant would be large enough to trigger inflation. Afterwards, as the universe decays into lower modes, the effective value of the cosmological constant would be getting smaller and smaller until it would reach the value $\Lambda_0$ that would be the currently observed value of the cosmological constant.

Another way to obtain a small effective value of the cosmological constant is to suppose a negative value for $\Lambda$. Then, $\Lambda^{\rm eff}_k \in [0, \Lambda_0]$ provided that $\Lambda$ is of the same order of $\Lambda_0$. However, it implies a strong fine tuning and it would mean that our universe is now in a state with a high value $r$ of the mode. This does not seems to be consistent.

\subsubsection{Case 2: $\lambda(a) \propto a$}

Let us now consider the coupling function
\begin{equation}
	\lambda^2(a) = - \frac{3 \pi M_P^2}{2} \alpha^2 a^2~,
\end{equation}
where $\alpha$ is a constant parameter. Now
\begin{equation}\label{FRQ05}
	\omega_k^2(a,\varphi) = \sigma^2 (H_{0}^2 a^4  - \alpha_k^2 a^2) + \sigma^2 H_1^2 a^4,
\end{equation}
with
\begin{equation}
	\alpha_k^2 = 1 + \alpha^2 \sin^2\frac{\pi k}{N} .
\end{equation}
Because the factor $\alpha_k^2$ in Eq. (\ref{FRQ05}), Eq. (\ref{WDW02}) with Eq. (\ref{FRQ05}) would not actually represent the quantum state of a closed universe. However, we can perform a scale factor transformation 
\begin{equation}
	\tilde{a}_k = \sqrt{\alpha_k} a ,
\end{equation}
in terms of which Eq. (\ref{WDW02}) turns out to be
\begin{equation}\label{WDW03}
	\ddot{\tilde{\phi}}_k + \frac{1}{\tilde{a}_k} \dot{\tilde{\phi}}_k - \frac{1}{\tilde{a}_k^2} \tilde{\phi}''_k + \omega_k^2(\tilde{a}_k,\varphi) \tilde{\phi}_k = 0 ,
\end{equation}
where now the dots stand for the derivative with respect to the transformed scale factor $\tilde{a}_k$, and
\begin{equation}\label{FRQ06}
	\omega_k^2(\tilde{a}_k,\varphi) = \sigma^2 (\tilde{H}_{0,k}^2 \tilde{a}_k^4  -  \tilde{a}_k^2) + \sigma^2 \tilde{H}_{1}^2 \tilde{a}_k^4,
\end{equation}
with
\begin{equation}
	\tilde{H}_{0,k}^2 = \frac{H_0^2}{\alpha_k^2} = 3 \frac{\Lambda_0}{(1+\alpha^2 \sin^2\frac{\pi k }{N})^2} = 3 \Lambda^{\rm eff}_k . 
\end{equation}
Eq. (\ref{WDW02}) with Eq. (\ref{FRQ06}) does actually represent the quantum state of a closed universe with an effective value of the cosmological constant given by $\Lambda_k^{\rm eff}$. 

For a value $-1 < \alpha^2 < 0$, the effective value of the cosmological constant satisfies, \mbox{$\Lambda_k^{\rm eff} \in [ \Lambda_0, \frac{\Lambda_0}{(1- |\alpha^2|)^2}] \rightarrow [\Lambda_0, \infty]$}, for  $|\alpha^2| \rightarrow 1$. It would imply that for large values of the mode $k$, at the onset of the universe, the effective value of the cosmological constant would be large enough to trigger inflation. However, as the state of the universe is decaying the effective value of the cosmological constant is decreasing. The current state of the universe would then correspond to a very small value of the mode~$k$.

\subsubsection{Case 3: constant value of $\lambda(a) \propto \sqrt{E_0}$}

In this case, Eq. (\ref{WDW02}) would represent the quantum state of a universe for which
\begin{equation}\label{FRQ07}
	\omega_k^2(a,\varphi) = \sigma^2 (H_{0}^2 a^4  -  a^2 + E_k) + \sigma^2 H_1^2 a^4,
\end{equation}
with
\begin{equation}
	E_k= E_0 \sin^2\frac{\pi k}{N} .
\end{equation}
Eq. (\ref{FRQ07}) is the frequency that arises in the third quantized model of a universe filled with a minimally coupled scalar field with mass $m$, like the field $\varphi$ considered in this paper, and another massless scalar field which is conformally coupled to the background space-time (see Refs. \cite{RP2013a, Gott&Li}). The conformally coupled masses scalar field can effectively mimic a radiation like field with an energy given by $E_k$ \cite{HH,OB&Mourao}. Therefore, the result of the interaction between universes would imply in the present case the appearance of a radiation like content of the universe that would be of order $E_0$ for large values of the mode $k$, and it would be decaying to the value $E_k \approx 0$ for small values of the mode ($E_k= 0$ for $k=0$). This effective content would imply the existence of a pre-inflationary stage of the universe that should have observable effects in the power spectrum of the CMB.

Let us us consider the flat branch of a De Sitter (or quasi-De Sitter, i.e. $\dot{H} \approx 0$) space-time. Then Eq. (\ref{FRQ07}) simplifies as,

\begin{equation}
 \omega_k^2(a,\varphi) = \sigma^2 (H^2 a^4  + E_k) ,
\end{equation}
with $H = 3 \Lambda$. The Friedmann equation turns out to be then
\begin{equation}
\dot{a} = \frac{\omega(a)}{a} = \frac{\sigma}{a} \sqrt{H^2 a^4 + E_k} ,
\end{equation}
with solutions given by
\begin{equation}\label{sf03}
a(t) = a_0 \sinh^\frac{1}{2} \theta ,
\end{equation}
where $\theta = 2 H\sigma t + \theta_0$, with $\theta_0 $ some constant to fit with the boundary conditions. At late times, the scale factor (\ref{sf03}) grows in an exponential way approaching to an exact De Sitter expansion. However, at the earliest epoch it shows a deviation from deSitter evolution that would have a strong influence in the lowest modes of the CMB power spectrum \cite{Bouhmadi2011, Scardigli2011, Bouhmadi2013}.

In Refs. \cite{Bouhmadi2013}, it is analyzed the effects that a preinflationary phase dominated by an energy density inspired by the generalized Chaplygin gas \cite{Bento2002} (see also Ref. \cite{OB-Duvvuri})
\begin{equation}\label{nfdw}
\rho = \left( \frac{B_1}{a^{\beta_1 (1+\alpha_1)}} + A_1 \right)^\frac{1}{1+\alpha_1} 
\end{equation}
has in the power spectrum of the CMB. Let us notice that Eq. (\ref{nfdw}) can reproduce a radiation dominated preinflationary stage of the universe, like the one studied in this paper, with a suitable choice of parameters: $\alpha_1 = 0$, $\beta_1 = 4$, $A_1 = \sigma^2 H^2$, and $B_1 = \sigma^2 E_k$. Then, the same procedure used in Ref. \cite{Bouhmadi2011} can be applied here. The results (see also Ref. \cite{Scardigli2011}) indicate that a radiation dominated preinflationary stage of the universe may alleviate the quadrupole anomaly of the CMB in a better way than ia matter dominated preinflationary stage. However, it is concluded that a greater departure from De Sitter space-time is needed during the preinflationary stage in order to better fit with the observational data.

\subsubsection{Case 4: Value of $\lambda(a) \propto a^{-1}$}

Let us now analyze the case where $\lambda(a) \propto \frac{1}{a}$. This case is particularly interesting  for at least two reasons: i) it shows also a preinflationary stage whose effects on the power spectrum of the CMB are expected to be stronger that those caused by a radiation dominated preinflationary stage \cite{RP2015}, and ii) the quantum effect of the interacting multiverse have no classical analogue so it can be considered  a distinguishable imprint of the multiverse on the cosmic observational data.

\begin{figure}[htbp]
\begin{center}
\includegraphics[width=8cm]{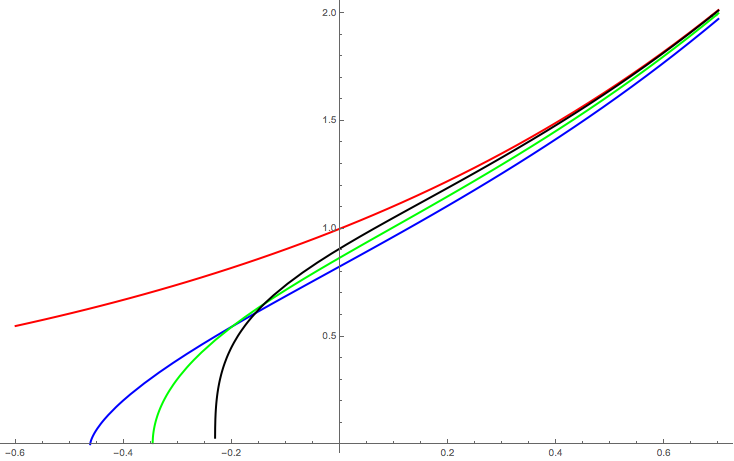}
\caption{Time evolution of the scale factor: i) flat De Sitter (red), ii) matter-dominated pre-inflationary stage (blue), iii) radiation-dominated pre-inflationary stage (green), and iv) interacting multiverse pre-inflationary stage (black).}
\label{fig5}
\end{center}
\end{figure}

Let us first point out that a term proportional to $a^{-2}$ in the frequency in Eq. (\ref{FRQ01}) arises also in the decomposition in partial waves of the wave function of a De Sitter space-time \cite{Garay2013}. Such a decomposition is equivalent to the interacting scheme presented here with a coupling function given by $\lambda(a) \propto a^{-1}$. It is therefore a pure quantum effect having no classical analogue. In both cases, it implies the appearance of a term proportional to $a^{-6}$ in the effective equation of the energy density. In contrast to the term $a^{-4}$ caused by the radiation dominated preinflationary stage, the departure from De Sitter space-time caused by the interacting multiverse is thus stronger. The effective Friedmann equation for the flat branch of the De Sitter universe turns out now to be given  by
\begin{equation}\label{sf04}
\dot{a} = \frac{\sigma}{a} \sqrt{H^2 a^4 + \frac{C_k}{a^2}} ,
\end{equation}
where $C_k$ is
\begin{equation}
	C_k= C_0 \sin^2\frac{\pi k}{N} .
\end{equation}
with $C_0$ is some constant parameter, and solutions 
\begin{equation}
a(t) = a_0 \sinh^\frac{1}{3} \theta ,
\end{equation}
and $\theta = 3 H\sigma t + \theta_0$, being $a_0$ and $\theta_0$ constants of integration. The departure form the De Sitter evolution lead by the term proportional to $a^{-2}$ in the Friedmann equation (\ref{sf04}) is stronger than the one produced by a radiation term (proportional to $a^0$) (see Fig. \ref{fig5}). Therefore, it is expected that its effects on the lowest modes of the power spectrum of the CMB would provide a better fit with the observed data \cite{RP2015}. It would provide an observational support to the model presented in this paper and to the whole multiverse proposal. Thus, it provides with distinguishable predictions that can be compared with observational data, making therefore falseable the whole multiverse theory.

\subsection{Double universe decay}

\begin{figure}[htbp]
\begin{center}
\includegraphics[width=8cm]{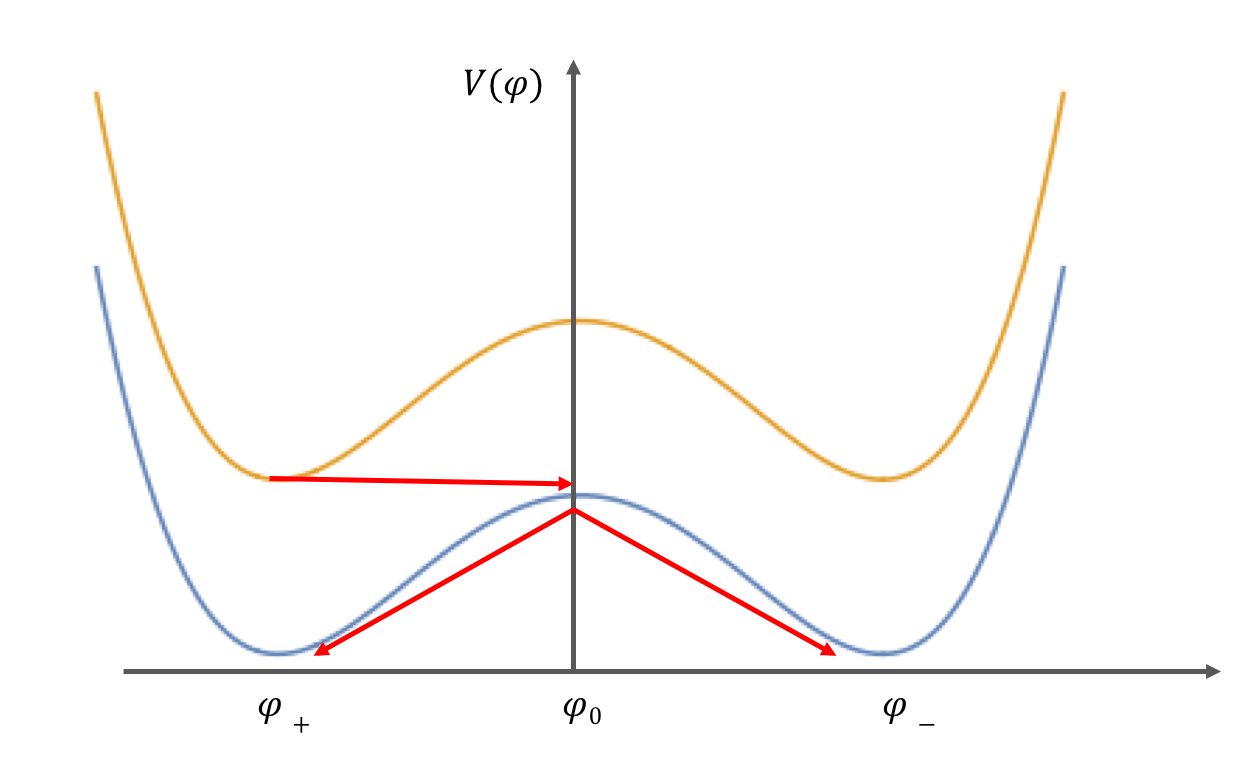}
\caption{The vacuum decay can occur in a two-bubble decay process through the metastable state at $V(\varphi_0)$.}
\label{fig4}
\end{center}
\end{figure}

The interacting picture developed in Sec. II opens up the possibility of new and interesting processes of vacuum decay such as the one depicted in Fig. \ref{fig4}. The decay between the vacuum state of the mode $r$ and those of the mode $k-1$ might occur through an intermediate vacuum decay into a metastable state given by the value $V(\varphi=0)$ of the mode $k-1$. This state would rapidly decay into the vacuum states of modes in a process that parallels those occuring in quantum optics where a two-photon state is generated through a metastable state (see, for instance, Ref. \cite{Yuen1975}), where the radiation field turns out to be described in terms of pairs of entangled photons (see Fig. \ref{fig4}).

In the case of the vacuum decay of the space-time, it would result into the generation of two bubbles of true vacuum (of the lower false vacua) whose quantum states would be entangled,
\begin{equation}\label{ES}
  \phi = \phi^+(\varphi_-) \phi^+(\varphi_+) \pm \phi^-(\varphi_-)\phi^-(\varphi_+) ,
\end{equation}
where $\phi^\pm(\varphi_\pm)$ are the expanding and contracting branches of Eq. (\ref{WFSC01}), with $V(\varphi_\pm) \approx V(\varphi_\pm) + \frac{1}{2} V''(\varphi_\pm) \varphi^2$.

The properties of the space-time inside the two entangled bubbles would  be the same at large scales for observers inhabiting therein. For instance, the effective value of the cosmological constant would be the same, given by $\Lambda_0$  and the effective mass scale of the scalar field would be in both bubbles given by $V''(\varphi_\pm) = m^2 (\frac{1}{\lambda_{\varphi}}-1)$, so the inner part of the two bubbles would be very similar at large scales.

If the two entangled bubbles would come out from a double instanton, like the one studied in Ref. \cite{RP2013a}, then, the quantum state of one of the bubbles would be given by the reduced density matrix that is obtained by tracing out the degrees of freedom of the partner bubble of the entangled pair, with a probability given by \cite{RP2013a}
\begin{equation}\label{P01}
  \Gamma/V \propto e^{- 2 I} ,
\end{equation}
with
\begin{equation}\label{I}
  I = \frac{a_+ H}{3} \left[ (a_+^2 + a_-^2) E(q) - 2 a_-^2 K(q)\right] ,
\end{equation}
where $K(q)$ and $E(a)$ are the complete elliptic integrals of first and second kind. We could follow Ref. \cite{RP2013a} to obtain the probability for the double Euclidean instanton given by the Euclidean solutions of Eq. (\ref{WDW01}) with the quartic potential (\ref{QV}). In general (see also Ref. \cite{RP2011b}), the resulted state is a thermal state that is indistinguishable from a classical mixture  so that observers inhabiting the bubbles would see the scalar field of their universes in a thermal state, being completely unaware of the entanglement properties of their bubbles. There is also a time reversal symmetry between the time variables of the entangled bubbles  \cite{RP2013a} so the regions inside the bubbles would present opposite symmetry assignments such as, for instance, baryon asymmetry or other discrete symmetries (see also Ref. \cite{OB&Mourao}) that would be the consequence of the global  symmetries of the entangled pair of bubbles. For an  observer external to the bubbles, the symmetry assignments and
 asymmetries would disappear when he/she would consider the properties of the whole entangled pair. The question then is whether this would fix some of the apparent asymmetries of our universe.

\section{Conclusions}

In this work we have examined the implications of an interacting multiverse and the issue of bubble nucleation of ``true" vacua in the universes filled by the ``false" vacuum at its genesis or later. The existence of a multiverse where the universes may interact opens up the door to a new scenario with important implications on the global structure of the universes. As a result of the interactions it appears a landscape structure where the universes are created with different effective values of the cosmological constant. Thus, there exist the possibility of quantum tunneling transitions between different universal states giving rise to new bubbles with the corresponding value of their vacuum states.

The interaction between universes and the generation of new bubbles is expected to be dominant only for small length scales of the parent space-time, where quantum effects of the space-time are significant. However, these newborn bubbles may expand and generate new bubbles in a self-reproducing process. 

The vacuum decay between the quantum states of two or more universes is expected to be cut off for large values of their scale factors. This condition imposes a restriction on the possible values of the coupling function in the interaction scheme presented in this paper. The possible cases of interest have been analysed and they all would have observable consequences in the global properties of each single universe. They could explain the very small value of the cosmological constant of our universe. However, a fine-tunning is always required making the proposal, in this respect, no better than others cosmological scenarios \cite{Alonso2012}. In the case that the coupling function would be a constant, it would appear the existence of a pre-inflationary stage in the evolution of the single universes. A pre-inflationary stage would have observable implications in the power spectrum of the CMB of the universe \cite{Bouhmadi2011}. Furthermore, it has been claimed that it would fi
 t some o
 f the anomalies that most models of inflation present with respect to the latest observational data provided by Planck. 

The interacting multiverse with a coupling function proportional to $a^{-1}$ provides us with a preinflationary stage of the universe whose departure from De Sitter expansion is stronger than the one caused by a radiation dominated preinflationary stage. This would be a pure quantum effect having no classical analogue so: i) it might hopefully alliviate the quadrupole anomaly of the CMB and, ii) it could provide distinguishable predictions of the multiverse that cannot be explained by any other known effect. Thus, the model of the multiverse presented in this paper entails distinguishable predictions that can be compared with current observational data being therefore falsifiable leading to implications that have not been considered so far.

\section*{Acknowledgments}

A. A-S. was partially supported by the Marsden grant {\it ``Topics in mathematical general relativity and theoretical cosmology"} administered by the Royal Society of New Zealand. The work of CB is supported by the FCT (Portugal) grant SFRH/BPD/62861/2009.


\begin{thebibliography}{99}

\bibitem{Susskind} L. Susskind, {\it The Anthropic Landscape of String Theory}, hep-th/0302219.

\bibitem{Bousso} R. Bousso and J. Polchinski, JHEP {\bf 0006}, 006 (2000).

\bibitem{Weinberg} S. Weinberg, {\it Living in the Multiverse}, hep-th/0511037.

\bibitem{Polchinski} J. Polchinski, {\it The Cosmological Constant and the String Landscape} hep-th/0603249.

\bibitem{Holman} R. Holman and L. Mersini-Houghton, {\it Why the Universe started from a low entropy state?} hep-th/0511102.

\bibitem{Ellis} G.F.R. Ellis, {\it On horizons and the cosmic landscape}, astro-ph/0603266.

\bibitem{Bertolami2008} O. Bertolami, Gen. Rel. Grav. {\bf 40},1891 (2008).

\bibitem{Linde} A. Linde, Phys. Lett. {\bf B200}, 272 (1988).

\bibitem{Alonso2012} A. Alonso-Serrano, C. Bastos, O. Bertolami and S. Robles-P\'erez, Phys. Lett. {\bf B719}, 200 (2013).

\bibitem{Everett} H. Everett, {\it Theory of the Universal Wavefunction}, Thesis, Princeton University, http://www.pbs.org/wgbh/nova/manyworlds/pdf/ dissertation.pdf (1956); Rev. Mod. Phys. {\bf 29}, 454 (1957).

\bibitem{Linde1986} A. Linde, Phys. Lett. {\bf B175}, 395 (1986).

\bibitem{Bousso&Susskind} R. Bousso and L. Susskind, Phys. Rev. {\bf D85}, 045007 (2012).

\bibitem{Nomura} Y. Nomura, Phys. Rev. {\bf D86}, 083505 (2012).

\bibitem{Bertolami2013} O. Bertolami and C. Herdeiro, Int. J. Mod. Phys. {\bf D22}, 1350068 (2013).

\bibitem{Strominger1990} A. Strominger, in {\it Quantum Cosmology and Baby Universes}, edited by S. Coleman, J.B. Hartle, T. Piran, and S. Weinberg (World Scientific, London, UK, 1990), vol. 7.

\bibitem{Alonso-Serrano:2014dsa} A.~Alonso-Serrano, L.~J.~Garay and G.~A.~Mena Marugan, Phys.\ Rev.\ D {\bf 90},  124074 (2014).

\bibitem{Ade2014} P. A. R. Ade et al., A and A {\bf 571}, 48 (2014).

\bibitem{Ijjas2013} A. Ijjas, P. J. Steinhard and A. Loeb, Phys. Lett. {\bf B723}, 261 (2013).

\bibitem{Aguirre:2007an} A.~Aguirre, M.~C.~Johnson and A.~Shomer, Phys.\ Rev.\ D {\bf 76}, 063509 (2007).

\bibitem{Hawking1988} S. Hawking, Phys. Rev. {\bf D37}, 904 (1988).

\bibitem{Garay2013} I. Garay and S. Robles-P\'erez, Int. J. Mod. Phys. {\bf D23}, 1450043 (2014).

\bibitem{Birrell1982} N. D. Birell and P. C. W. Davies, {\it Quantum fields in curved space} (Cambridge University Press, Cambridge, UK, 1982).

\bibitem{Hartle1990} J. B. Hartle, in {\it Quantum Cosmology and Baby Universes}, edited by S. Coleman, J.B. Hartle, T. Piran, and S. Weinberg (World Scientific, London, UK, 1990), vol. 7.

\bibitem{Coleman1} S. Coleman, Phys. Rev. {\bf D15}, 2929 (1977); {\bf D16}, 1248 (1977).

\bibitem{Coleman2} C. Callan and S. Coleman, Phys. Rev. {\bf D16}, 1762 (1977).

\bibitem{Coleman1980} S. Coleman and F. De Luccia, Phys. Rev. {\bf D21}, 3305 (1980).

\bibitem{Mottola} E. Mattola and A. Lapedes, Phys. Rev. {\bf D29}, 773 (1984).

\bibitem{RP2011b} S. Robles-P\'erez and P. F. Gonz\'alez-D\'iaz (2011) arXiv: 1111.4128.

\bibitem{RP2013a} S. Robles-P\'erez, J. Phys. Conf. Ser. {\bf 410}, 012133 (2013).

\bibitem{Gott&Li} J. R. Gott  and L. X. Li, Phys.Rev.  {\bf D58}, 023501 (1998).

\bibitem{HH} J. B. Hartle and S.W. Hawking, Phys. Rev. {\bf D28}, 2960 (1983).

\bibitem{OB&Mourao} O. Bertolami and J. M. Mour\~ao, Class. Quant. Grav. {\bf 8}, 1271 (1991). 

\bibitem{Bouhmadi2013} M. Bouhmadi-Lopez, P. Chen, Y. Huang and Y. Lin, Phys. Rev. {\bf D87}, 103513 (2013).

\bibitem{Bento2002} M. C. Bento, O. Bertolami and A. A. Sen, Phys. Rev. {\bf D66}, 043507 (2002). 

\bibitem{OB-Duvvuri} O. Bertolami and V. Duvvuri, Phys. Lett. {\bf B640}, 121 (2006). 

\bibitem{Bouhmadi2011} M. Bouhmadi-Lopez, P. Chen and Y. Liu (2011), Phys. Rev. {\bf D84}, 023505 (2011).

\bibitem{Scardigli2011} F. Scardigli, C. Gruber and P. Chen (2011), Phys. Rev. {\bf D83}, 063507 (2011).


\bibitem{RP2015} S. Robles-P\'erez, M. Bouhmadi-Lopez and J. Marto, \emph{Effects of the multiverse on the power spectrum of the CMB}, in preparation. 

\bibitem{Yuen1975} H. P. Yuen, Phys. Lett. {\bf A51}, 1 (1975).







\end{thebibliography}

\end{document}